# A Statistical Approach to Modeling Indian Classical Music Performance

[1]Soubhik Chakraborty*, [2]Sandeep Singh Solanki, [3]Sayan Roy, [4]Shivee Chauhan, [5]Sanjaya Shankar Tripathy and [6]Kartik Mahto

[1]*Department of Applied Mathematics, BIT Mesra, Ranchi-835215, India*
[2, 3, 4, 5, 6]*Department of Electronics and Communication Engineering, BIT Mesra, Ranchi-835215, India*

Email: soubhikc@yahoo.co.in (S.Chakraborty)
sssolanki@bitmesra.ac.in (S.S. Solanki)
roysayan_2002@rediffmail.com (S. Roy)
shivee.chauhan@gmail.com (S. Chauhan)
sstripathy@bitmesra.ac.in (S.S.Tripathy)
kartik_mahto@rediffmail.com

*S. Chakraborty is the corresponding author (phone: +919835471223)

*Words make you think a thought. Music makes you feel a feeling. A song makes you feel a thought.*                  -------- E. Y. Harburg (1898-1981)

## Abstract

A raga is a melodic structure with fixed notes and a set of rules characterizing a certain mood endorsed through performance. By a vadi swar is meant that note which plays the most significant role in expressing the raga. A samvadi swar similarly is the second most significant note. However, the determination of their significance has an element of subjectivity and hence we are motivated to find some truths through an objective analysis. The paper proposes a probabilistic method of note detection and demonstrates how the relative frequency (relative number of occurrences of the pitch) of the more important notes stabilize far more quickly than that of others. In addition, a count for distinct transitory and similar looking non-transitory (fundamental) frequency movements (but possibly embedding distinct emotions!) between the notes is also taken depicting the varnalankars or musical ornaments decorating the notes and note sequences as rendered by the artist. They reflect certain structural properties of the ragas. Several case studies are presented.

The paper is organized as follows. Section 1 is the introduction. In section 2 we attempt to model a Pilu performance using multinomial distribution. Section 3 has more statistical analysis where transitory and non-transitory frequency movements are carefully analyzed. In section 4 we substantiate our claims with two more studies in ragas Yaman and Kirwani. Finally section 5 is reserved for conclusion.

**1. Introduction**

A major strength of statistics lies in modeling. Tremendous progress in computer technology and the consequent availability of scores (musical notation) in digitized version has popularized modeling a musical performance greatly in the recent past. This is particularly true in Western Classical music (where scores are fixed). However, from the point of view of the listener or the critic, there is no unique way of analyzing a performance . This means a realistic analysis of musical performance cannot be cannot be purely causal or deterministic whence statistics and probability are likely to play an important role [1]. Coming to Indian classical music, where scores are not fixed (the artist decides extempore what to perform on the stage itself) we can only analyze a recorded performance. In this context, in an earlier work, we showed how statistics can distinguish performances of two different ragas that use the same notes [2]. A good text on Indian Classical music is Raghava R. Menon [3]. For technical terms, see Amrita Priyamvada [4]. Readers knowing western music will benefit from [5].

The present paper opens with a statistical case study of performance of raga Pilu using the only notes *Sa, Sudh Re, Komal Ga, Sudh Ma, Pa, Komal Dha and Sudh Ni* played on a scale changer harmonium(see the remark below) by the first author (*Sa* set to natural C) and recorded in the Laptop at 44.100 KHz, 16 bit mono, 86 kb/sec mode. This raga also permits the notes Sudh Ga, Sudh Dha and Komal Ni (which we have avoided for a comparison with another raga called Kirwani introduced from the South Indian classical music to the North Indian classical music) and is one of restless nature (*chanchal prakriti*), the appropriate time for playing it being between 12 O'Clock noon to 3 P. M. See [6] and [3]. However, like Bhairavi it has a pleasant effect if rendered at any other time different from the scheduled. It is rendered more by instrumentalists and in Thumris and Tappas by vocalists. Many experts are of the view that Pilu is created by mixing four ragas, namely, Bhairavi, Bhimpalashree, Gauri and Khamaj [6]. In our analysis of the Pilu performance, special emphasis is given on modeling and we raise questions such as "What is the probability of the next note to be a *Vadi Swar* (most important note in the raga)?" or "In what proportions the raga notes are arriving?" and seek answers through multinomial modeling (see appendix), reserving quasi multinomial model as the model of choice in case of failure of the former to meet our expectations.

**Remark (what is a scale changer harmonium and why is it necessary?)**: In a scale changer harmonium, such as the one used by the first author, the meaning of the phrase "Sa set to natural C" is that (i) the first white reed of the middle octave which normally represents C has not been shifted right or left to represent any other note or in other words we can say that the reed which represents C is at its natural position and that (ii) the performer has kept his Sa at this white reed itself. A scale changer harmonium has the charming facility of "shifting the scale" which assists in accompaniment. Let us take an example. Suppose the main performer (say a vocalist) has his Sa at C# (C-sharp) but the harmonium player is accustomed to playing with Sa taken at the aforesaid white reed which represents C. In this case the harmonium player would simply shift the white reed one place to the right so that the same reed would now represent C# rather than C and then accompany the main performer without difficulty. With this shift, all reeds are shifted by one place to the right (and we say there is a change of scale). Without a scale changer harmonium, the accompanist would be forced to take Sa at the first black reed of the middle octave which represents C#. Since he is not comfortable in moving his fingers by taking Sa here, the quality of accompaniment may be affected adversely. In a scale changer harmonium, the reeds (or keys) are not fixed to the reed-board. Rather they are fixed to another board and the instrument is fixed to a big tape. When the tape is moved, the reeds also change their places accordingly. Nowadays scale changer harmoniums are used in all major concerts of Indian music especially classical music. Although built in India, a harmonium is generally western tuned with imported German reeds.

**2. Modeling the Pilu performance statistically: multinomial or quasi multinomial?**

Since the instrument used is a scale changer harmonium, identifying the notes was done easily adopting the following strategy. First reeds (keys of the harmonium) corresponding to all the twelve notes of each of the three octaves were played sequentially (Sa, Komal Re, Sudh Re, Komal Ga, Sudh Ga, Sudh Ma, Tibra Ma, Pa, Komal Dha, Sudh Dha, Komal Ni, Sudh Ni) and recorded with Sa taken at natural C (see the remark above). Their frequencies (meaning fundamental frequencies) fluctuating with small variability each about a different near horizontal line ("horizontal" due to stay on each note) were extracted using Solo Explorer 1.0 software (see also section 3 and 4) and saved in a text file. Next the means and standard deviations of the fundamental frequencies of the notes were estimated from the values in the text file.

We provide for the benefit of the reader the database for the middle octave only which is used more often than the other two octaves. The fundamental frequencies of corresponding notes for the first and third octaves can be approximated by dividing and multiplying the mean values by 2 respectively. The standard deviation values are nearly the same in all the three octaves and hence omitted.

Table 1: Database for the middle octave

| Note | Mean fundamental frequency (Hz) | Standard Deviation (Hz) |
|---|---|---|
| Sa | 243.2661 | 0.4485 |
| Komal Re | 257.6023 | 0.1556 |
| Sudh Re | 272.3826 | 0.0503 |
| Komal Ga | 287.6051 | 0.2155 |
| Sudh Ga | 305.2415 | 0.1805 |
| Sudh Ma | 323.1398 | 0.3172 |
| Tibra Ma | 342.2261 | 0.2205 |
| Pa | 362.4957 | 0.4241 |
| Komal Dha | 384.4443 | 0.1316 |
| Sudh Dha | 407.6329 | 0.2227 |
| Komal Ni | 432.5978 | 0.1387 |
| Sudh Ni | 457.4805 | 0.3030 |
| Sa (next octave) | 484.9670 | 1.8249 |

Next from Chebyshev's inequality we know that a random variable X, irrespective of whether it is discrete or continuous satisfies the inequality
$P\{|X-E(X)| < k\, S.D.(X)\} \geq 1-1/k^2$. Here E(X) gives the mathematical expectation of X (which gives its population mean) and k is a constant. S.D. (X) gives the standard deviation of X and P is the probability function. The symbol | | stands for absolute value. In other words, we can say, by breaking the modulus that the inequality gives a lower probability bound to an interval in which the random variable would lie, i.e.,
$P\{E(X) - kS.D.(X) < X < E(X) + kS.D.(X)\} \geq 1-1/k^2$. For a formal proof of this inequality, see [7]. *The beauty of the inequality is that there is no distributional assumption on X*. Thus X need not follow any specific probability distribution. Now, applying the popular "6-sigma" limits, that is taking k=6, we get,
$P\{E(X) - 6S.D.(X) < X < E(X) + 6S.D.(X)\} \geq 35/36$.

**Remark: An important application of Chebyshev's inequality and six-sigma limits can be found in statistical quality control. See [8].**

*In our study, the random variable X represents the fundamental frequency, of which pitch is a function (see "pitch profile" later), characterizing a note*. Chebyshev's inequality gives a range in which the fundamental frequency for each note in a specific octave would practically lie with a large confidence coefficient (probability) not less than 35/36 (97.22%). We shall establish later that the overall arrival of notes in the sequence is indeed random. Having already estimated E(X) and S. D. (X) for the notes of the middle octave in table 1 (and halving or doubling the mean values for the notes of the other octaves) we could easily detect the sequence of notes and generate tables 2.1 to 2.4. *It is important to mention here that we could detect more notes using Chebyshev's inequality through the text file generated by the software (Solo Explorer 1.0 whose analysis is given later) than what this software could itself detect and exhibit graphically!* This is one of the strengths of this paper. Of course one had to be a bit careful here as the fundamental

frequencies have to come in succession for a while for endorsement as a note else it is merely a "glide".

Table 2.1: Frequency distribution (no. of occurrences of pitch) of notes in Pilu performance of one minute duration:-

| Note | Frequency (no. of occurrences of pitch) | Relative frequency = frequency/115 |
|---|---|---|
| Sa | 30 | 0.26087 |
| Sudh Re | 22 | 0.191304 |
| Komal Ga | 21 | 0.182609 |
| Sudh Ma | 08 | 0.069565 |
| Pa | 11 | 0.095652 |
| Komal Dha | 06 | 0.052174 |
| Sudh Ni | 17 | 0.147826 |
| Total | 115 | 1 |

Our hypothesis is that the overall distribution (from table 2.1) to be applicable to be applicable for smaller parts of the piece (Pilu performance on harmonium) as well.

Table 2.2: Frequency (no. of occurrences of pitch) distribution of notes in Pilu performance for the first 30 seconds:-

| Note | Observed Frequency (no. of occurrences of pitch) | Expected Frequency = relative frequency(table 2.1)*55 |
|---|---|---|
| Sa | 16 | 14.34785 |
| Sudh Re | 15 | 10.52172 |
| Komal Ga | 10 | 10.043495 |
| Sudh Ma | 01 | 3.826075 |
| Pa | 03 | 5.26086 |
| Komal Dha | 01 | 2.86957 |
| Sudh Ni | 09 | 8.13043 |
| Total | 55 | |

Pooling the frequencies of Sudh Ma, Pa, Komal dha and Sudh Ni (no theoretical cell frequency can be less than 5 as otherwise Chi-Square which is a continuous probability distribution loses it character of continuity; for more on Chi-Square see [7]), we get four classes for which Chi-square calculated for 3 degrees of freedom is 3.941011. Chi-Square for goodness of fit is calculated by the formula (observed fr – expected fr)/expected fr summed over all classes after pooling, if need be, and this follows Chi-Square distribution with degrees of freedom one subtracted from the number of classes after pooling.

Remark: The term degrees of freedom refers to the number of independent pieces of information (one from each variable; here one from each class) and one degree of freedom is lost because of the restriction that the sum of observed frequencies must equal the sum of expected frequencies. A plus point of this Chi-Square ("goodness of fit") test is that it is non parametric meaning "distribution-free", i.e., can be tried for any population. A negative point is that how we are forming the classes can affect its value marginally. This limitation is well known among statisticians and for a skilled personnel, this is never a major handicap and should not affect the result of the test in general except for the borderline cases for which one can cross check the results with a Kolmogorov-Smirnov test. The first author of this paper, who has contributed both as a performer and an analyst, and who is a statistician by profession feels it important to mention that the classes should be pooled in a way that no theoretical cell frequency is less than 5 and at the same time we get the maximum number of classes that we can after pooling. Only adjacent classes should be pooled.

Table Chi-Square for 3 degrees of freedom at 5% level of significance is 7.81 which means the probability of Chi-Square to exceed 7.81 if the hypothesis that the fit is good, (i.e. observed and expected frequencies are not significantly different) is true, is only 5%. So if the observed Chi Square exceeds this value we would take it that the fit is bad or in other words the observed and expected frequencies are significantly differing at 5% level of significance. Here calculated Chi-Square (3.941011) is less than table Chi Square and therefore insignificant at 5% level of significance.

Table 2.3: Frequency (no. of occurrences of pitch) distribution of notes in Pilu performance for the middle 30 seconds (20 sec to 50 sec):-

| Note | Observed Frequency (no. of occurrences of pitch) | Expected Frequency = relative frequency(table 2.1)*72 |
|---|---|---|
| Sa | 17 | 18.78264 |
| Sudh Re | 08 | 13.773888 |
| Komal Ga | 13 | 13.147848 |
| Sudh Ma | 06 | 5.00868 |
| Pa | 10 | 6.886944 |
| Komal Dha | 06 | 3.756528 |
| Sudh Ni | 12 | 10.643472 |
| Total | 72 | |

Chi Square at 6 df for above = 5.707324 (df= degrees of freedom)
Table Chi-Square for 6 degrees of freedom at 5% level of significance is 12.59.
Calculated Chi-Square is less than table Chi Square and therefore insignificant at 5% level of significane. Here no pooling of classes is needed (all cell frequencies exceed 5)

Table 2.4: Frequency (no. of occurrences of pitch) distribution of notes in Pilu performance for the last 30 seconds:-

| Note | Observed Frequency (no. of occurrences of pitch) | Expected Frequency = relative frequency(table 2.1)*60 |
|---|---|---|
| Sa | 14 | 15.6522 |
| Sudh Re | 07 | 11.47824 |
| Komal Ga | 11 | 10.95654 |
| Sudh Ma | 07 | 4.1739 |
| Pa | 08 | 5.73912 |
| Komal Dha | 05 | 3.13044 |
| Sudh Ni | 08 | 8.86956 |
| Total | 60 | |

Arguing as in table 2.2, pooling the frequencies of Sudh Ma and Pa in one class and also pooling the frequencies of Komal dha and Sudh Ni in another class, we get 5 classes for which Chi-square calculated for 4 degrees of freedom is 4.615536.

Table Chi-Square for 4 degrees of freedom at 5% level of significance = 9.49

Note: We emphasize again that the term "frequency" in tables 2.1-2.4 and also in tables 4.1-4.3 refers to number of occurrences of the pitch characterizing a note.

We make the following interesting interpretations:-

1. The relative frequencies of the overall one minute performance is best reflected in the performance of the middle 30 seconds. This observation is, however, performer dependent.

2. *The Chi-square values for the first, middle and last 30 seconds are all insignificant*. This means the overall probabilities of notes are not varying significantly from trial to trial but this does not imply that the probabilities are equal for each individual note as well! Remember classes were pooled. However, on the whole, we may accept the multinomial model at 5% level at least in the present performance, if we can endorse independence of notes in the note sequence, and would caution the reader that overall equality is a weaker than individual equality so that the reader is free to experiment with a Quasi multinomial model for better results. See also point 4.

3. In particular, the relative frequencies of Sa, Komal Ga and Sudh Ni are very stable for all the three groups compared to the overall performance. *We strongly feel this finding worth reporting immediately, given that Sa is always the base*

> *note while Komal Ga and Sudh Ni are theoretically known to be the Vadi and Samvadi swars (most important and next most important notes) respectively in the raga Pilu. The experimental results thus endorses the theory and in the bargain gives a clue for detecting objectively the top two important notes in the raga. That said, we do propose to apply the technique in ragas such as Bageshree and Shankara in future where there is a difference of opinion regarding which swars should be Vadi and Samvadi (see also section 4).*

4. Writing 1 for Sa, 2 for Sudh Re, 3 for Komal Ga etc. and using run test for randomness(see [7] and appendix), we verified that the overall note sequence in entirety of the Pilu performance may be taken to be random(the total number of runs coming out to be 57 against the expected 58.5). However, this does not guarantee independence for every pair of notes. For example, the cumulative frequencies of Vadi and Samvadi swars clearly showed a pattern as detailed later. Overall randomness implies overall independence of notes (which does not prove independence in pairs!) and is an assumption of multinomial distribution. See also appendix.

Remark: If the performer is a novice, one must keep the option of a wrong note also as a possibility (with corresponding probability to be estimated from actual performance similarly) to make the possible cases exhaustive.

### 3. More Statistical Analysis: analysis of transitory and non-transitory frequency movements

In addition to the study on modeling a performance, a count for distinct transitory and similar looking non-transitory frequency movements (but possibly embedding distinct emotions!) between the notes is also taken. It is held that the emotions of the raga are contained in these frequency (and hence pitch) movements apart from the note sequences [9]. In the rendition of ragas in Indian music, not only the notes and note sequences but *how they are rendered* characterize the raga. The "how" part has to do with the transitory and even non-transitory frequency movements between the notes. For example, the sequence {Komal Ga, Sudh Re, Sa} of raga Bhimpalashree would be rendered differently if realized as the last three notes of the sequence {Sudh Ma, Komal Ga, Sudh Re, Sa} in raga Bageshree; the difference being caught in the aforesaid frequency movements. There is also a concept of alankar in Indian music meaning ornament (of course in a musical sense!). The shastras have categorized alankars into Varnalankar and Shabdalankar[10]. The varnas include sthayi (stay on a note), arohi (ascent or upward movement), awarohi (descent or downward movement) and sanchari (mixture of upward and downward movement). This classification of alankars relate not only to the structural aspect of the raga, but also the raga performance in that " all the extempore variations that a performer created during a performance within the raga and tala limits could be termed as alankar, because these variations embellished and enhanced the beauty of the raga, the tala and the composition."[9] The case of other pictures like hats and valleys is under investigation.

Similarly a careful study of note sequences revealed those sequences as are typical of the raga Pilu. Prominent among them are {Sudh Ni, Sa, Re, Komal Ga}, {Komal Ga, Re, Sa, Sudh Ni}, {Komal Ga, Re, Sa, Sudh Ni, Sa} where the Sudh Ni is of the previous(first) octave. The combination {Pa, Komal Dha, Sudh Ni, Sa}, which is acceptable in another raga Kirwani that uses the same notes as used in the present Pilu performance, was also successfully detected. This is a typical combination used in both the ragas which are otherwise different. Note that use of the notes Sudh Ga, Sudh Dha and Komal Ni is also permissible in the raga Pilu which makes it more colourful. We have refrained from doing this here as in a subsequent paper we plan to compare this performance with one with Kirwani. See also section 4.

Here is a typical sample of Pilu from Solo Explorer 1.0 software, which is a pitch to midi converter and music transcriber. The horizontal lines correspond to (fundamental) frequencies while the vertical lines depict the arrival of a note (in Western notation) with onset times. The software has the provision of generating a text file of onset times and fundamental frequencies (which we omit for shortage of space):-

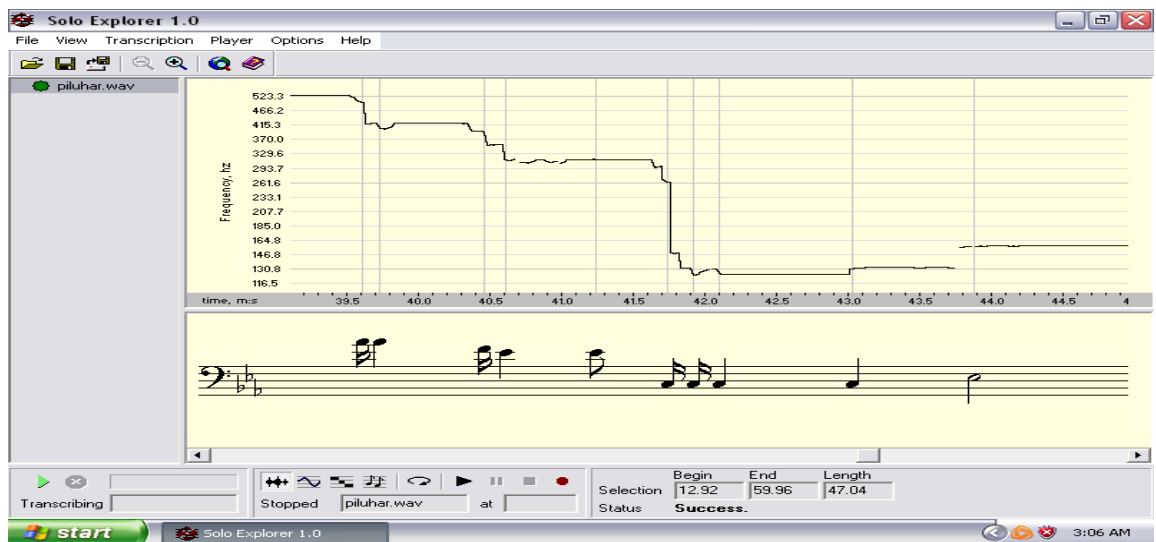

Fig 1: Solo Explorer 1.0 graph of Pilu performance sample

We next show some pictures which may help identify a specific transitory frequency movement. A transition here means a shift from one note to another either in discrete (in jumps) or in continuous mode; a non-transitory movement refers to a horizontal line with some tremor indicating a stay on a certain note; as mentioned earlier a stay can also contain emotion and and stays on different notes or on the same note on different occasions can contain different emotions; hence although all such stays would graphically look alike as innocent looking horizontal lines with some tremor, we should not neglect them.

Archive of Cornell University e-library; arXiv:0809.3214v1[cs.SD][stat.AP].
http://arxiv.org/abs/0809.3214

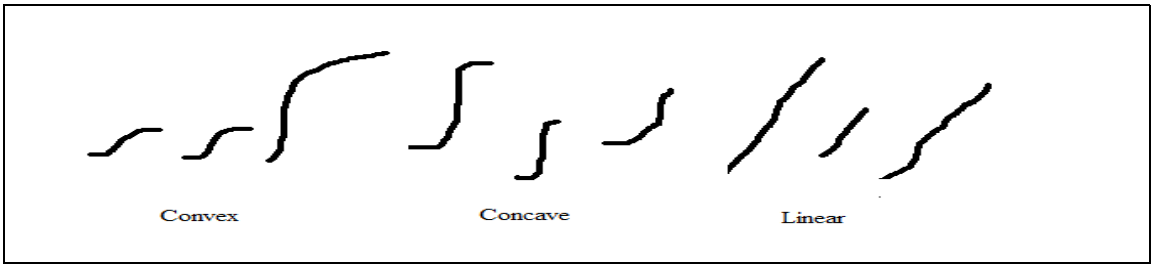

Fig. 2.1 Example of Rising Transition

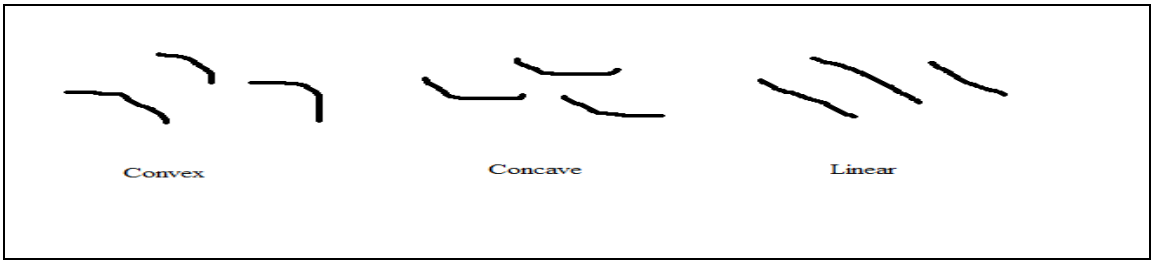

Fig. 2.2 Example of falling transition

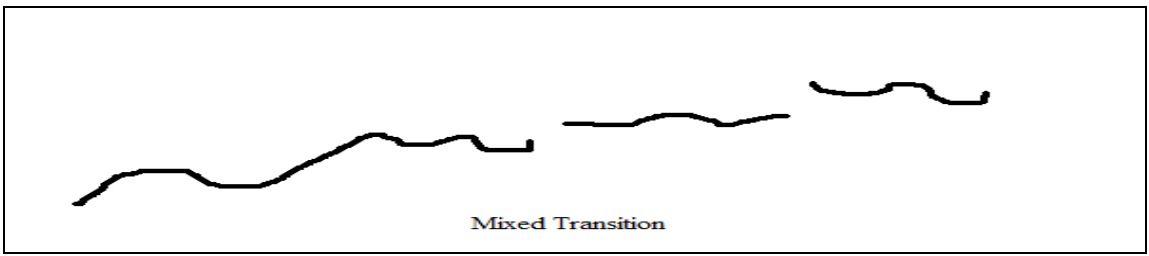

Fig. 2.3 Example of mixed transition

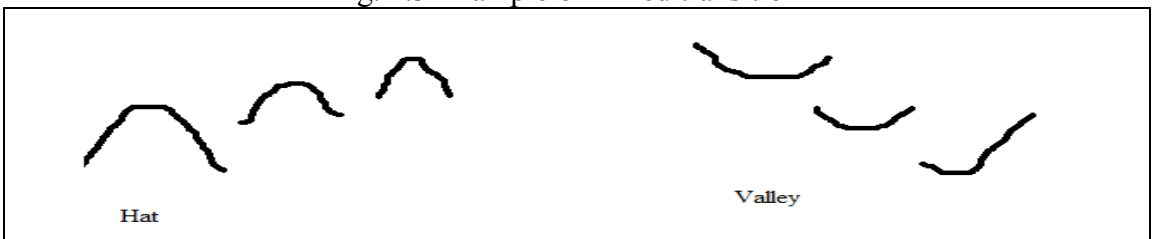

Fig. 2.4 Example of hat and valley

These figures which are also given in [9] and which were drawn independently by the fourth author helped us greatly in generating tables 3.1 and 3.2. The reader by looking at these figures can understand how we must have generated tables 3.1 and 3.2.

Table 3.1 Transitory and non-transitory frequency movements for Pilu from Solo Explorer 1.0 software

| Rising transitions | | | Falling Transitions | | | Mixed transitions | No Transition |
|---|---|---|---|---|---|---|---|
| 18 | | | 24 | | | 15 | 34 |
| Convex | Concave | Linear | Convex | Concave | Linear | | |
| 03 | 04 | 11 | 08 | 06 | 10 | | |

Table 3.2 More transitory frequency movements for Pilu:-

| Hats shaped as "^" | | | Valleys shaped as "v" | | |
|---|---|---|---|---|---|
| 06 | | | 11 | | |
| Positively skewed | Negatively skewed | Symmetric | Positively skewed | Negatively skewed | Symmetric |
| 01 | 03 | 02 | 03 | 02 | 06 |
| Low | Moderate | High | Shallow | Moderate | Deep |
| 03 | 02 | 01 | 05 | 04 | 02 |

Interpretation:-

1. Falling transitions are more, which is expected as the raga Pilu has a tendency to descend (*Awarohi varna*). This finding is valuable structurally for the raga though it is endorsed through a performance (which means even a novice can make an intelligent guess that Pilu is probably a raga of Awarohi varna).

2. The number of valleys are about twice the number of hats. This is not very easy to interpret this finding but we suspect the cause could again be some emotional manifestation of what is given in the earlier point. As a matter of fact, what kind of emotions are embedded in specific frequency movements demand a very serious study involving expert artists. We hope ITC Sangeet Research Academy, Kolkata, India which has the necessary infrastructure and the expert musicians will take up the challenge sometimes in the near future. Final interpretation can be given only thereafter.

3. Staying on a note at different instances can also be aesthetic in different ways though graphically we are absolutely clueless about the innocent looking near horizontal lines with some tremor. Nevertheless there is no harm in keeping track of where they occurred dominantly for a longer duration till the arrival of the next note which in our case turned out to be at onset times 2.5, 8.62, 11.55, 12.93, 17.22, 19.74, 22.72, 25.17, 29.10, 32.95, 36.24, 42.11, 43.89, 49.44, 53.93. The corresponding notes of such prominent stay are respectively Sa, Komal Ga, Komal Ga, Sa, Sa, Ni, Pa, Sa, Sa, Sa, Sa, Ni, Komal Ga, Sa, Sa. In Indian Classical music, Sa and Pa are obvious notes for stay (exception Pa of raga Bageshree) while Komal Ga and Sudh Ni are the most important and next most important notes of Pilu as mentioned earlier. Pa in spite of being a stay note has been used only once for long duration. Although this is performer dependent and not unexpected in a one minute recording let us not forget that Pa unlike Sa is not a base note (where we come back again and again) and is neither a Vadi nor a Samvadi swar.

A graph of onset times of notes of longer duration in MS Excel 2003 package depicted an approximate linear pattern (fig. 3.1). [The legend "series 1" does not mean anything and is created by default]. There are 15 notes of longer stay (long horizontal line with some tremor) out of 34 horizontal lines in all (34 non transitory movements) with tremor. The remaining 19 horizontal lines were of short length. Here "long" and "short" were decided subjectively by looking at the graphs; we could fix that stay of a certain number of seconds minimum would be taken as stay but "a certain number of seconds" is itself subjective! There is not much point arguing over this issue; we relied on our intuition remembering that there is no unique way of explaining a musical performance (yes!).

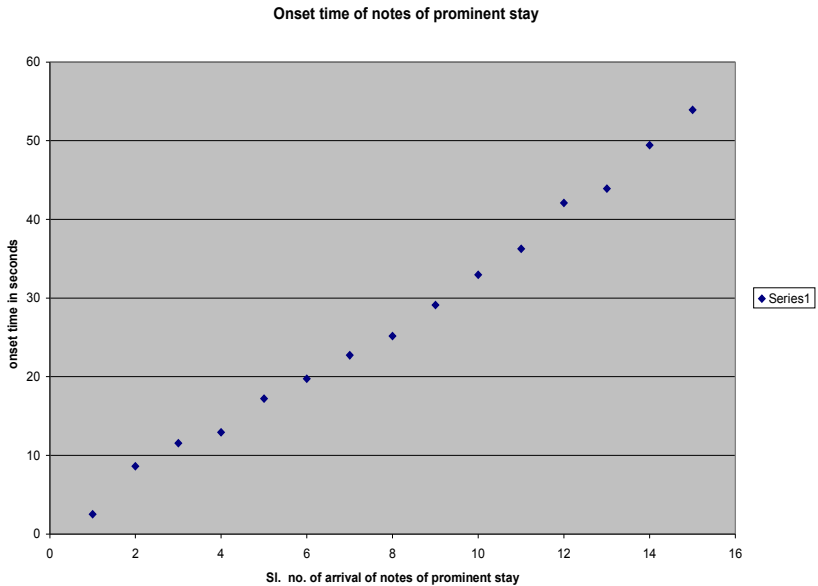

Fig. 3.1 onset graph for Pilu only for notes of longer duration (subjective)

A graph of onset time of all the notes is given next using MATLAB (notice the clustering of notes as they were played with a fast tempo very similar to a *taan)*:-

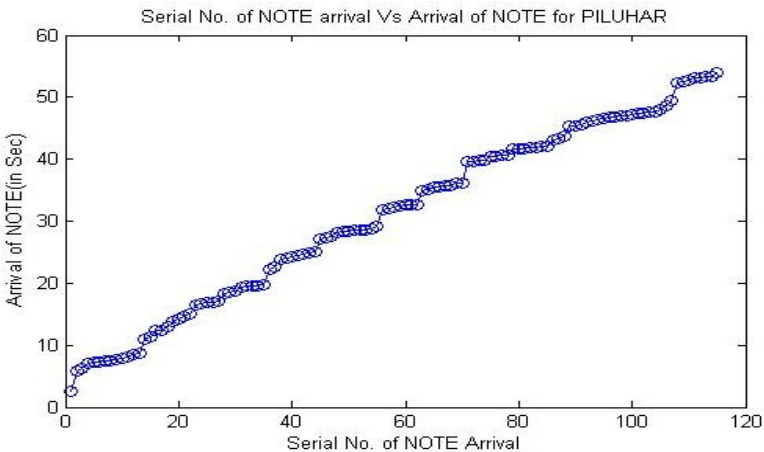

Fig 3.2 Onset graph of all the notes detected from Pilu performance

Interpretation: The onset graph of only the stay notes in fig. 3.1 gives a more or less linear pattern depicting that the performer stayed on a note fairly regularly. On the other hand, clustering of notes in Fig. 3.2 ocurred because the performer played with a fast tempo very similar to *taans* and that even this has been done at regular intervals! See also section 4 for a more interesting inter-onset interval graph for studying the rhythm in note arrival.

**Pitch profile**
We next give the pitch profile which is of interest because *pitch (p) is the perceived fundamental frequency (f) of a sound.* It is the "sensation" of this frequency. *Shrillness or hoarseness of a note is determined by pitch only.* The higher the pitch, the more shrill the note. Music theorists sometimes represent pitch using a numerical scale based on the logarithms of fundamental frequency. For example one can adopt the formula $p= 69 +12 * \log_2 [f /(440 Hz)]$. This creates a linear pitch space in which octaves have size 12, semitons (the distance between adjacent keys on the piano keyboard) have size 1,and A440 is assigned the number 69. Distance in this space corresponds to musical distance as measured in psychological measurements and understood by musicians. The system is flexible enough to accommodate microtones not found on the standard piano keyboard. The pitch representation here is MIDI. The highest pitch was recorded as 83 and the lowest 58.

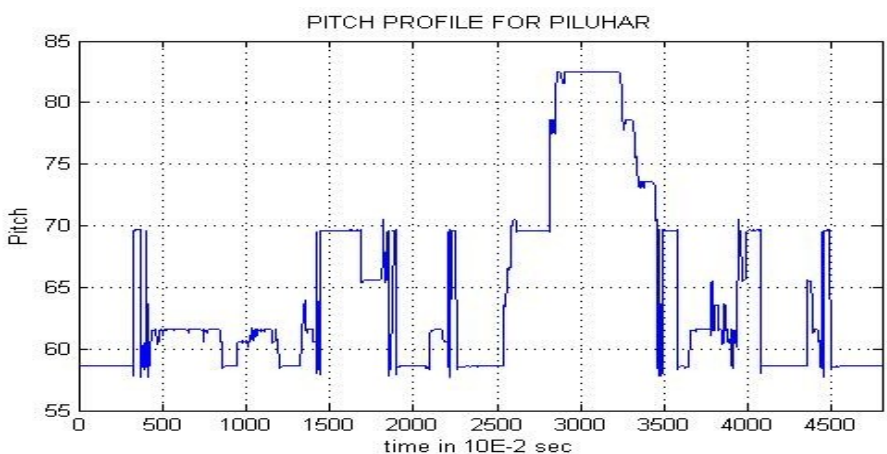
Fig. 3.3 Pitch profile for Pilu performance

**How are frequencies of Vadi and Samvadi swars related? (discovering a pattern in the occurrences of the top two important notes with respect to each other)**

For the Pilu performance, the variances and covariance of Vadi and Samvadi swars are respectively Var (Vadi)=17.16524, Var(Samvadi)=14.48695 and covariance(Vadi, Samvadi)= -3.104351(see appendix for the formulae).
Hence correlation coefficient = -0.19686. This is low enough to rule out at least a linear relationship though other relationship can be present.

Further investigation revealed the following table:-

Table 3.3 Frequencies (f) and cumulative frequencies (cf) for Vadi Swar
(Komal Ga) and Samvadi Swar (Sudh Ni) of Pilu Performance
**Frequency means number of occurrences of pitch**

| Seconds | f of Vadi Swar | cf of Vadi swar | f of Samvadi swar | cf of samvadi swar |
|---|---|---|---|---|
| 0-10 | 04 | 04 | 02 | 02 |
| 10-20 | 04 | 08 | 02 | 04 |
| 20-30 | 02 | 10 | 05 | 09 |
| 30-40 | 00 | 10 | 04 | 13 |
| 40-50 | 10 | 20 | 03 | 16 |
| 50-60 | 01 | 21 | 01 | 17 |

**Cumulative frequency is of less than type here. Cumulative frequency (less than type) of a class is defined as the number of observations less than the upper limit of the class**.

A plot of the frequencies of Vadi and Samvadi swars clearly depict a non-linear pattern. In fact, fit to a polynomial of degree four is found excellent (fig. 3.4). Although notes are independent overall, the top two important notes seem to be heavily dependent. Is this not interesting? Drawing cumulative frequency (cf) curves (cf against the class intervals) is left as an exercise for the reader. There will be two cf curves, one for the Vadi swar and the other for the Samvadi swar.

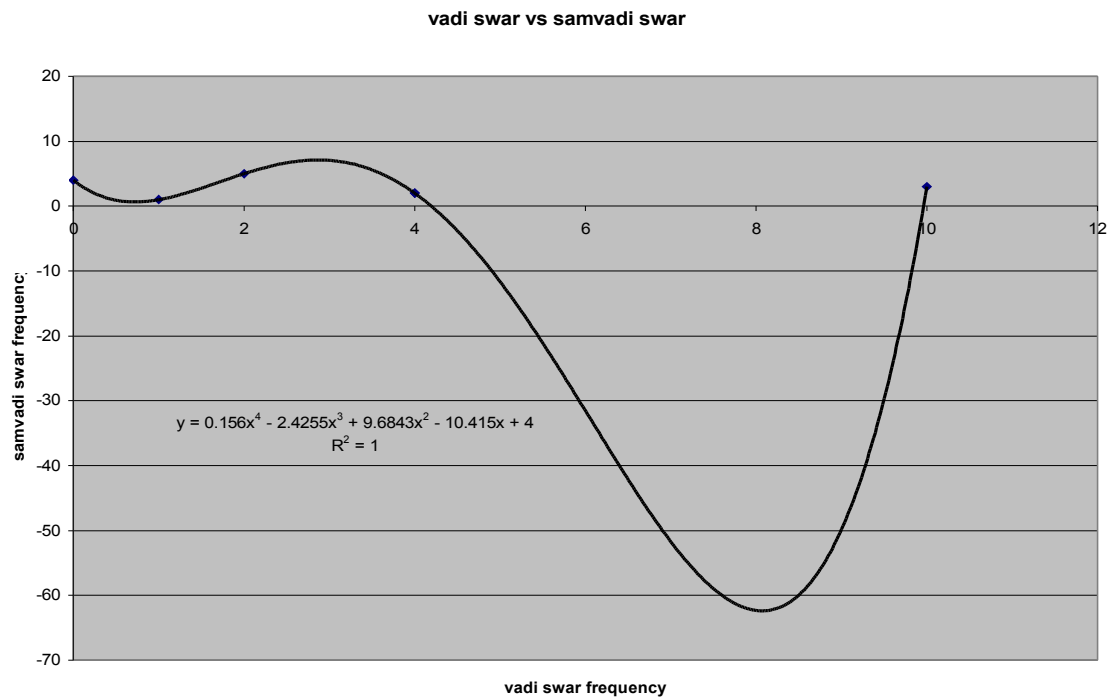

Fig 3.4: Frequency plots for the Vadi and Samvadi swars

## 4. Substantiating our claims with two more studies

Empirical definition of probability implies *probability is relative frequency in the long run*. It is striking that even in a one minute recording the probability of Vadi and Samvadi swars are remarkably stable over the three short pieces of 30 seconds each as compared to the overall performance. A recording of longer duration would perhaps stabilize the relative frequency of the less important notes of the raga also! There are four important lessons to be learnt from this - (i) *it seems that the most important note in a raga is not necessarily one which is used more often (as this could well be a stay note or Nyas swar like Sudh Re in Yaman) but one whose relative frequency, apart from being high (though not necessarily highest), stabilizes faster in a short period of time*. In a private communication to the first author, Prof. Alan Marsden (Lancaster University, UK) made an interesting remark that, since the scientist himself is the performer here, it is possible that the data is unconsciously created to suit what the scientist has in mind! He suggested "recording a performer who is not otherwise engaged in the research is the only way to be sure of getting data which is certain not to be pre-determined to any degree by the researcher's pre-existing ideas". We are not psychologists and on this issue therefore would not like to debate further. But there is some hope for all of us. Wife of the first author, a trained vocalist (North Indian Classical) and a D. Mus. from Banaras Hindu University, has rendered a recording on our request, again of one minute duration, in raga Yaman in which the Vadi swar Sudh Ga has relative frequency 0.127072 for the whole minute and 0.129310 for the middle 30 seconds (accidentally again the best representative), 0.181818 for the first 30 seconds and 0.104000 for the last 30 seconds. The relative frequency of the Samvadi Swar Sudh Ni is found to be 0.303867 for the whole minute and 0.310344 for the middle 30 seconds, 0.218181 and 0.336000 for the first 30 and last 30 sec respectively. She is not a scientist and asks innocent questions such as "*What has statistics to do with music*?" to her statistician husband or sometimes, in a more challenging tone, "*Are you trying to mean my voice has so much tremor as your 'nonsense' graphs are showing*?" Explaining to her why pitch, even in a perfectly steady-sounding voice, is actually never steady was quite a task! In order to substantiate matters, we are providing below features of the raga Yaman and a table of relative frequency of all the notes over the whole minute as well as in first 30 sec, middle 30 sec and last 30 sec (table 3.4). A detailed analysis of her recording is under review [11].

Features of raga: *Yaman*

*Thaat* (a specific way of grouping ragas): Kalyan
*Aroh* (ascent):  *N*  R G, m P, D, N, **S**
*Awaroh* (descent): **S** N D, P, mG,RS
*Jati*: *Sampooorna-Sampoorna* (7 distinct notes allowed in ascent and 7 in descent; however, since there are only 12 possible notes, some notes have to be common between ascent and descent)
*Vadi Swar* (most important note): G
*Samvadi Swar* (second most important note): N
Prakriti (nature): Restful
Pakad (catch): *N* R G R, *N* R S
Nyas Swar (Stay notes): R G N P

Time of rendition: First phase of night (6 PM to 9 PM)

**Remark**: This raga was created by Amir Khusru.

**Abbreviations**

The letters S, R, G, M, P, D and N stand for Sa(always Sudh), Sudh Re, Sudh Ga, Sudh Ma, Pa(always Sudh), Sudh Dha and Sudh Ni respectively. The letters r, g, m, d, n represent Komal Re, Komal Ga, Tibra Ma, Komal Dha and Komal Ni respectively. A note in Normal type indicates that it belongs to middle octave; if in italics it is implied that the note belongs to the octave just lower than the middle octave while a bold type indicates it belongs to the octave just higher than the middle octave

Table 3.4 Relative frequency (no. of occurrence of pitch) of the notes of Yaman

| Note | Whole minute | Ist 30 sec | middle 30 sec | last 30 sec |
| --- | --- | --- | --- | --- |
| Sa | 0.220994 | 0.181818 | 0.241379 | 0.240000 |
| Sudh Re | 0.149171 | 0.163636 | 0.146551 | 0.144000 |
| Sudh Ga | 0.127072 | 0.181818 | 0.129310 | 0.104000 |
| Tibra Ma | 0.066298 | 0.109090 | 0.060344 | 0.048000 |
| Pa | 0.044198 | 0.072727 | 0.034482 | 0.032000 |
| Sudh Dha | 0.088397 | 0.072727 | 0.077586 | 0.096000 |
| Sudh Ni | 0.303867 | 0.218181 | 0.310344 | 0.336000 |

It is interesting to note that the Samvadi Swar Sudh Ni, which is also getting stable, has a higher relative frequency compared to the Vadi Swar and this is because it is also a stay note like the Vadi Swar. For that matter, even Sudh Re which is nether Vadi nor Samvadi has high relative frequency next only to that of Sa and it is again due to its being a stay note. Again an important note cannot have low relative frequency as well. *We conclude that high relative frequency is only a necessary condition for being important but never sufficient. The sufficient condition is what is stated in (i).*

It follows from claim (i) that (ii) *controversies which still exist among Indian musicians as to which notes should be Vadi and Samvadi in certain ragas like Bageshree and Shankara can perhaps be settled now*. In Bageshree, the Vadi and Samvadi Swars are Sudh Ma and Sa according to some experts whereas others say they are Sudh Dha and Komal Ga. Similarly in raga Shankara they should be Sudh Ga and Sudh Ni according to some while others say they should be Pa and Sa respectively. The other two lessons are that (iii) if one has a recording of longer duration it still makes sense to sample a small subset from it for analysis – the advantage with a longer recording however is that there can be several such samples from different positions reflecting perhaps different moods of the performer in rendering the raga and (iv) we need to have a recording of sufficiently longer duration to get more precise values of probabilities of the less important notes (as they take some time to settle). Nevertheless multinomial modeling demands only an overall stability of probability of all notes collectively (this is weaker than saying it holds for each individual note) and this we have established through Chi-Square tests at 5% level of significance**.**

IMPORTANT: This paper raises the question "How do I decide objectively which swar is Vadi swar?" which is quite different from the question why the Vadi swar is so called. It is so called as it plays the most important role in expressing the raga. This means, apart from elaborating the mood characterizing the raga, it also tells whether the raga is *purvanga pradhan* (first half more important) or *uttaranga pradhan* (second half more important) according as it is one of the notes from Sa to Pa or from Ma to **Sa.** As Ma and Pa fall in both the halves thus created, expert guidance is needed to decide the more important half in case the Vadi swar turns out to be one of these two notes. Also the Vadi swar gives us a rough idea of the timing of the raga's rendition. For a *purvanga pradhan* raga, such as Pilu, the timing broadly is 12 O' clock to 12 P.M. For an *uttaranga pradhan raga*, such as Bhairav, the time period is somewhere between 12 PM to 12 O'clock. [6].

**Results of run test**: The total number of runs (57) came very close to the expected (58.5). This implies an overall independence of note arrival in the note sequence (see appendix). However, cumulative frequencies of Vadi and Samvadi swars are well represented by a polynomial of degree four for data points over first 20, first 30….first 60 seconds implying how overall independence does not mean independence in pairs. Since for multinomial modeling, it is overall independence of the notes in the sequence rather than their individual independence in pairs that was wanted hence, in summary, we may accept the multinomial model for the present performance at 5% level of significance (the model could be better if the duration of the recording had been longer; this would allow the relative frequency of even the less important notes to stabilize) but we do emphasize that a quasi multinomial model could still be better and that we must keep such an option open for other situations. *Detecting the Vadi and Samvadi Swars in a situation when the relative frequencies are all unstable and quasi-multinomial distribution prevails is an open research problem.*

Metric analysis is done next with an inter onset interval time graph (time between successive onsets of notes). Peaks about a horizontal line in the graph correspond to rhythm (depicting equal time between successive onsets. We provide an interesting graph of inter-onset interval for Pilu (Fig. 3.5) in MATLAB and this is followed by a comparison of Pilu and Kirwani performances for the modeling part only.

Archive of Cornell University e-library; arXiv:0809.3214v1[cs.SD][stat.AP].
http://arxiv.org/abs/0809.3214

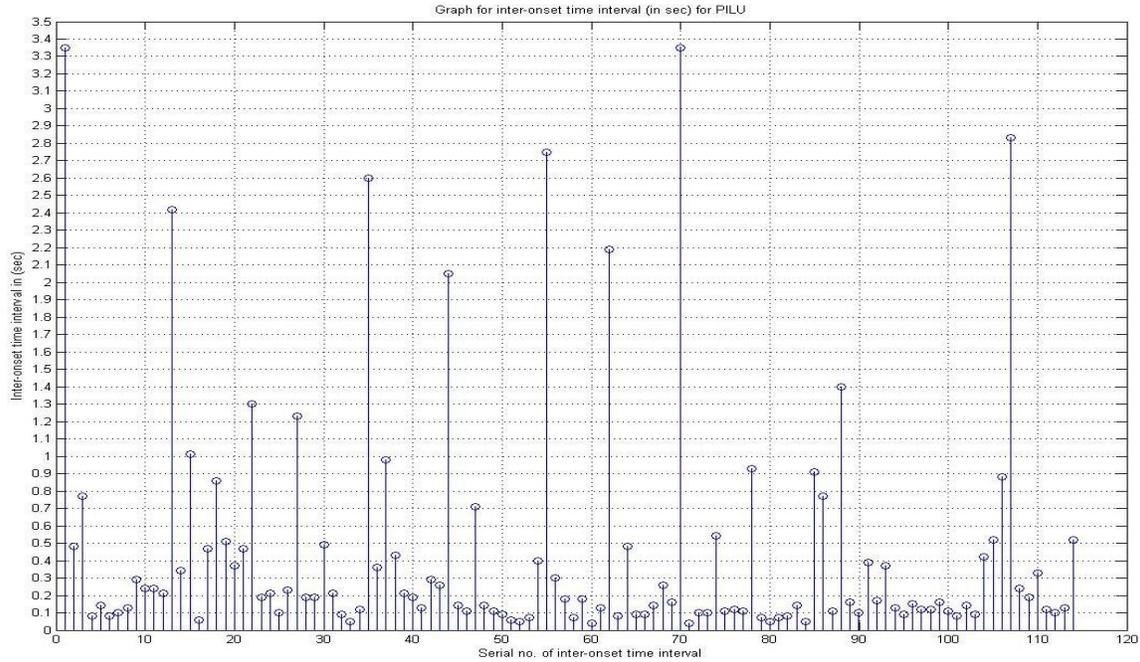

**Fig 3.5 Inter-onset interval graph in MATLAB for Pilu**

**Comparing relative frequency with another raga Kirwani which uses the same notes as used in the Pilu performance:-**

The following frequency distribution (table 4.1) was obtained in another one minute recording of Kirwani using the same notes as in the Pilu recording.

Table 4.1: Frequency distribution of notes in Kirwani performance of one minute duration:-

| Note      | Frequency | Relative frequency = frequency/82 |
|-----------|-----------|-----------------------------------|
| Sa        | 09        | 0.109756                          |
| Sudh Re   | 11        | 0.134146                          |
| Komal Ga  | 09        | 0.109756                          |
| Sudh Ma   | 08        | 0.097561                          |
| Pa        | 21        | 0.256098                          |
| Komal Dha | 16        | 0.195122                          |
| Sudh Ni   | 08        | 0.097561                          |
| Total     | 82        | 1                                 |

A moment's reflection upon comparison of table 4.1 with table 2.1 reveals a marked difference in the relative frequencies for Sa, Pa and Komal Dha. The relative frequencies of Sudh Ma are the closest in both followed by those of Sudh Re and Sudh Ni. The relative frequencies of Komal Ga are different though not so pronounced.

Table 4.2: Frequency table segmented

| Notes | First 30 sec | | middle 30 sec (20 sec to 50 sec) | | last 30 sec | |
|---|---|---|---|---|---|---|
| | fr | relative fr | fr | relative fr | fr | relative fr |
| Sa | 03 | 0.061224 | 08 | 0.123077 | 06 | 0.181818 |
| Re | 06 | 0.122449 | 09 | 0.138461 | 05 | 0.151515 |
| Komal Ga | 05 | 0.102041 | 07 | 0.107692 | 04 | 0.121212 |
| Ma | 05 | 0.102041 | 07 | 0.107692 | 03 | 0.090910 |
| Pa | 14 | 0.285714 | 15 | 0.230769 | 07 | 0.212121 |
| Komal Dha | 13 | 0.265306 | 12 | 0.184615 | 03 | 0.090910 |
| Ni | 03 | 0.061224 | 07 | 0.107692 | 05 | 0.151515 |
| Total | 49 | | 65 | | 33 | |

Once again the relative frequency of the notes for the overall performance is best reflected in the middle 30 seconds only. We again maintain that this is performer dependent and a rule must not be formed though as mentioned earlier, vocal performance by wife of the first author in another raga has produced a similar finding. The frequencies of Komal Ga and Sudh Ma are the most stable. **It seems Komal Ga is the most important note in both the ragas but Sudh Ma is more important in Kirwani than in Pilu. Pa is also more important in Kirwani than in Pilu as a stay note**. In ragas such as Kirwani and Charukeshi, introduced to North Indian Classical from South Indian Classical, rules are not so strict regarding Vadi-Samvadi. Hence this comparison deserves a special attention.

A table of expected frequencies is supplied next.

Table 4.3: Table of expected frequencies for the three segments

| Notes | First 30 sec | | middle 30 sec (20 sec to 50 sec) | | last 30 sec | |
|---|---|---|---|---|---|---|
| | fr | expected fr rel fr(table 5)*49 | fr | expected fr rel fr(table 5)*65 | fr | expected fr rel fr(table 5)*33 |
| Sa | 03 | 5.38 | 08 | 7.13 | 06 | 3.62 |
| Re | 06 | 6.57 | 09 | 8.72 | 05 | 4.43 |
| Komal Ga | 05 | 5.38 | 07 | 7.13 | 04 | 3.62 |
| Ma | 05 | 4.78 | 07 | 6.34 | 03 | 3.22 |
| Pa | 14 | 12.55 | 15 | 16.65 | 07 | 8.45 |
| Komal Dha | 13 | 9.56 | 12 | 12.68 | 03 | 6.44 |
| Ni | 03 | 4.78 | 07 | 6.34 | 05 | 3.22 |
| Total | 49 | | 65 | | 33 | |

One can calculate the chi squares as before by pooling classes wherever necessary and endorse the validity of the multinomial model (details omitted). Note that pooling the classes is a bit subjective since the only restriction is that no theoretical cell frequency should be less than 5. Depending on how we are pooling, the final value of Chi-square can change marginally. Fig. 3.6 gives a multiple bar diagram depicting comparative number of occurrences of the notes of Pilu and Kirwani.

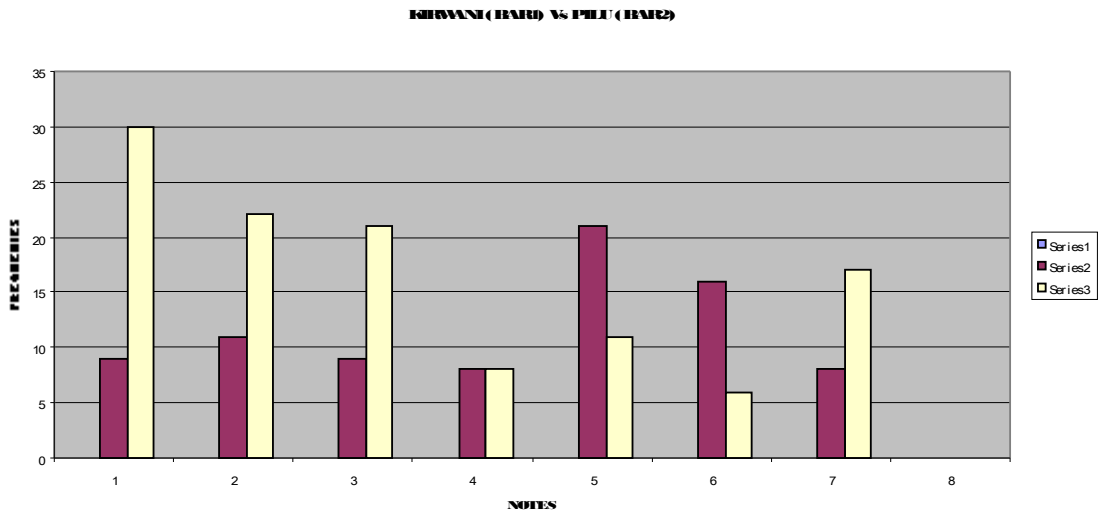

Fig. 3.6 Comparative frequencies of notes of Pilu and Kirwani

In the X axis, 1-7 respectively imply Sa, Sudh Re, Komal Ga, Sudh Ma, Pa, Komal Dha and Sudh Ni. (ignore the legends series 1- 3 which the package created by default).

For the reader's interest, the following are the transitory and non-transitory pitch movements for Kirwani (compare with Pilu):-

Rising transition 16 (Convex 4; Concave 5; Linear 7)

Falling transition 24 (Convex 5; Concave 9; Linear 10)

Mixed transition 12

No transition 19

Hats 14 {(+vely skewed 7; -vely skewed 5; symmetric 2); (Low 5; moderate 4; deep 5)}

Valleys 16 {(+vely skewed 3; -vely skewed 2; symmetric 11); (shallow 4; moderate 5; deep 7)}

The rising and falling transitions are similar in both the Pilu and Kirwani performances, with falling transitions surpassing rising transitions. There is interestingly less non-transitory movements in Kirwani indicating a more restless nature.

**Section 5  Conclusion**

As a final comment, finding the fundamental frequency of a complex acoustic signal is an arduous task. See [12] and the references cited therein. In particular [13] deserves a mention. In the context of Indian music see [14]. Although we have relied heavily on the Solo Explorer 1.0 software to do this for us, our signal processing team verified the results with some ingenious MATLAB programs. *We feel it important to mention here that the Solo Explorer software has to be set to the appropriate instrument whose signal it is supposed to receive for transcription. The option list unfortunately did not contain harmonium though it did contain, interestingly, several instruments used in Indian music such as Shehnai and Sitar). It was only after an intuitive but painstaking trial and error with other instruments from the option list such as piano, harmonica, organ etc. we discovered that if it this software is set to harmonica and fed with harmonium signals, it gives best results as cross checked from MATLAB programs.*

Another strength of our paper lies in successfully demonstrating that in an instrument with fixed reeds such as a harmonium or a piano, the Chebyshev's inequality can be used quite effectively for note detection purpose. Since it is the raga that we intend to study and not the instrument, such an unconventional simple statistical approach is a novelty in itself and demands immediate attention from musicians, music critics and statisticians (who know music). The first author was playing a scale changer harmonium with Sa set to natural C. His wife also sang with Sa at C only. The database for note detection was also created with Sa taken at C. In a vocal performance, in contrast, Sa could be elsewhere depending on the artist's convenience and it is another challenge then to first determine where the Sa is! Another problem is detection of microtones (Shruti swars) like a note which is sharper (of higher pitch) than C but flatter (of lower pitch) than C#. The relevant papers of interest are [15] and [16]. At the time of submitting this paper, we are successful in procuring two commercial recordings of scale changer harmonium though both are in the form of a jumble, that is, with other accompanying instruments. Our signal processing team is trying to develop a software that can "filter out" a particular instrument's signals for analysis. In addition, we are also trying our best to gather some studio recordings where each individual instrument can be independently recorded by a separate microphone.

As a future work, we also intend to see whether incorporating some additional notes such as Sudh Ga, Sudh Dha and Komal Ni in Pilu (which we avoided here) significantly contribute to the melody of the raga or not. If it does, we can argue that purity of the raga and melodic properties of notes are two different issues and one need not imply the other. To prove this point, in a subsequent paper, the first author promises to play and record the raga Pilu for a longer duration (involving the additional notes), corrupt the raga (in a musical sense!) and yet make it more colourful and hence increase the melodic properties of the notes. And of course we do promise to settle the controversy in ragas like Bageshree and Shankara where we have difference of opinion in Vadi-Samvadi selection.

[Concluded]

**Acknowledgement**

We wholeheartedly thank Prof. Alan Marsden and two anonymous referees for making several valuable comments to improve the paper. One referee pointed out "It is perhaps worth looking at research by Krumhansl (for example 'Cognitive Foundations of Musical Pitch', 1990, Oxford University Press) since she did research on pitch stability and tonal hierarchies in Western music. It might be nice to compare these with your findings." This is reserved as a rewarding future work. The first author, however, has an objection to the other referee's inclination of proposing Sudh Dha to be used at the expense of Komal Dha in Pilu. According to Late Suresh Chakraborty, grandfather of the first author and well known musicologist of his time (formerly associated with AIR Kolkata), Pilu is close to Bhairavi in its pure form. And Bhairavi in its pure form uses Komal Dha and not Sudh Dha. Sudh Dha in Mishra Pilu is acceptable but should be used only for colourfulness as in Mishra Bhairavi.

Finally, the first author thanks his wife Purnima for the Yaman recording and Dr. (Ms.) Vanamala Parvatkar, formerly Head, Faculty of Performing Arts (Vocal), Banaras Hindu University for a very fruitful telephonic conversation related to the present findings.

**References**

1. J. Beran and G. Mazzola, Analyzing Musical Structure and Performance -A Statistical Approach, Statistical Science, 1999, vol. 14, no. 1, p. 47-79.

2. S. Chakraborty, S. S. Solanki, S. Roy, S. S. Tripathy and G. Mazzola, A Statistical Comparison of Performance of Two Ragas(Dhuns) that Use the Same Notes, Proceedings of the International Symposium on Frontiers of Speech and Music, Kolkata, Feb 20-21, 2008, 167-171

3. R. R. Menon, Indian Music: The Magic of the Raga, Manohar Publishers and Distributors, 2007

4. A. Priyamvada, Encyclopaedia of Indian Music, Eastern Book Corporation, 2007.

5  C. S. Jones, Indian Classical Music, Tuning and Ragas,
    http://cnx.org/content/m12459/1.6/

6. D. Dutta, Sangeet Tattwa (Pratham Khanda), Brati Prakashani, 5$^{th}$ ed, 2006(in Bengali)

7. S. C. Gupta and V. K. Kapoor, Fundamentals of Mathematical Statistics, Sultan Chand And Sons, New Delhi, 8$^{th}$ ed. 1983

8. B. C. Gupta and H. F. Walker, Statistical Quality Control for the Six Sigma Green Belt, American Society for Quality, 2007

Archive of Cornell University e-library; arXiv:0809.3214v1[cs.SD][stat.AP].
http://arxiv.org/abs/0809.3214

9. R. Sengupta, N. Dey, D. Nag and A. K. Dutta, Extraction and Relevance of Transitory Pitch Movements in Hindustani Music, Proceedings of the National Symposium on Accoustics, New Delhi, 2006

10. http://www.itcsra.org/alankar/alankar.html

11. S. Chakraborty, R. Ranganayakulu, S. Chauhan and S. S. Solanki, A Probabilistic method of ranking musical notes, International Journal of Mathematical Modeling, Simulation and Applications(submitted)

12. L. R. Rabiner, M. J. Cheng, A. E. Rosenberg, and C. A. McGonegal, A comparative performance study of several pitch detection algorithms, IEEE Trans. Acoust., Speech, Signal Processing, vol. ASSP-24, no. 5, pp. 299-418, 1976

13. A. P. Klapuri, Multiple fundamental frequency estimation based on harmonicity and spectral smoothness, IEEE Trans. Speech and Audio Proc. 11(6), 804-816, 2003

14. A. K. Dutta, Pitch Analysis of Recorded Vocal Performances in Hindustani Music: Evidence of a Personal Scale, J. Acoust. Soc. India, Vol. XXV, 1997

15. R. Sengupta, Automatic tonic (Sa) detection algorithm in Indian Classical Vocal music, National Symposium on Accoustics, Bangalore, 2005

16. R. Sengupta, Automatic Extraction of Swaras and Srutis from Kheyal Renditions, J. Acoust. Soc. India, vol. 30, 2002

# Appendix

**Multinomial distribution**

Consider n independent trials being performed. In each trial the result can be any one of k mutually exclusive and exhaustive outcomes e1, e2…ek, with respective probabilities p1, p2…pk. These probabilities are assumed fixed from trial to trial. Under this set up, the probability that out of n trials performed, e1 occurs x1 times, e2 occurs x2 times ….ek occurs xk times is given by the well known multinomial law[4]

$\{n!/(x_1! \, x_2!...x_k!)\} p_1^{x_1} p_2^{x_2} \ldots p_k^{x_k}$ where each xi is a whole number in the range 0 to n subject to the obvious restriction on the xi's, namely, x1+x2+…..+xk=n.

E(xi)=n(pi) and Var(xi)=n(pi)(1-pi), i=1, 2…….k. Cov (xi, xj) = -n(pi)(pj)
One can then calculate the correlation coefficient between xi and xj as
Cov(xi, xj)/√[Var(xi)*Var(xj)]

If the probabilities vary we shall get a *quasi multinomial distribution*.

**Run Test of Randomness**

Let $x_1, x_2…x_n$ be n observations arranged in the order of arrival or occurrence. Calculate their median. Now examine each observation and write L if it is less than the median and M if it is more. If equal, write either L or M. You will get a sequence such as LLMLLLMMMMLLMMMMM…etc. Count the number of runs = U. **A run is a sequence of letters of one type preceded and/or followed by letters of another type**. For example, LLMLLLMMMM has four runs (counting one each for LL, M, LLL, MMMM).

Under the null hypothesis that observations are random, U is a random variable with
$E(U)=(n+2)/2$ and $Var(U)=(n/4)[(n-2)/(n-1)]$

For large n, the statistic $Z = [U-E(U)][+\sqrt{Var(U)}]$ follows standard normal distribution.
If the absolute value of $Z >= 1.96$, the null hypothesis, that the observations are random, will be rejected at 5% level of significance otherwise it may be accepted [4].

In the present analysis we had n=115 (total number of notes) and U=57 leading to
Z = - 0.280987, insignificant clearly to endorse overall randomness at 5% level of significance. As mentioned earlier, overall independence does not imply independence in pairs. If the reader knows probability theory he can recollect that the property that does guarantee independence for every pair and in fact independence for any subset taken from the whole is called mutual independence [7]. But we never established this here!